\documentclass[sigconf]{acmart}
\usepackage[T1]{fontenc}
\AtBeginDocument{%
  \providecommand\BibTeX{{%
    \normalfont B\kern-0.5em{\scshape i\kern-0.25em b}\kern-0.8em\TeX}}}

\acmConference[AISec '23]{Co-located with ACM-CCS'23}{\today}{Copenhagen}
\acmPrice{15.00}
\acmISBN{978-1-4503-XXXX-X/18/06}

\usepackage{balance} 

\copyrightyear{2023} 
\acmYear{2023} 
\setcopyright{acmlicensed}\acmConference[AISec '23]{Proceedings of the 16th ACM Workshop on Artificial Intelligence and Security}{November 30, 2023}{Copenhagen, Denmark}
\acmBooktitle{Proceedings of the 16th ACM Workshop on Artificial Intelligence and Security (AISec '23), November 30, 2023, Copenhagen, Denmark}
\acmPrice{15.00}
\acmDOI{10.1145/3605764.3623909}
\acmISBN{979-8-4007-0260-0/23/11}

\begin{document}

\title{Dictionary Attack on IMU-based Gait Authentication}

\author{Rajesh Kumar}
\email{rajesh.kumar@bucknell.edu}
\orcid{0000-0001-7467-5762}
\affiliation{%
  \institution{Bucknell University}
  \city{Lewisburg}
  \country{USA}}

\author{Can Isik}
\email{cisik@syr.edu}
\orcid{0000-0003-2894-4540}
\affiliation{%
  \institution{Syracuse University}
  \streetaddress{College Place}
  \city{Syracuse}
  \country{USA}}

\author{Chilukuri Krishna Mohan}
\email{ckmohan@syr.edu}
\affiliation{%
  \institution{Syracuse University}
  \streetaddress{College Place}
  \city{Syracuse}
  \country{USA}}
\renewcommand{\shortauthors}{Rajesh Kumar, Can Isik, \& Chilukuri Krishna Mohan}






\begin{abstract}
We present a novel adversarial model for authentication systems that use gait patterns recorded by the inertial measurement unit (IMU) built into smartphones. The attack idea is inspired by and named after the concept of a dictionary attack on knowledge (PIN or password) based authentication systems. In particular, this work investigates whether it is possible to build a dictionary of \textit{IMUGait} patterns and use it to launch an attack or find an imitator who can actively reproduce \textit{IMUGait} patterns that match the target's \textit{IMUGait} pattern. Nine physically and demographically diverse individuals walked at various levels of four predefined controllable and adaptable gait factors (speed, step length, step width, and thigh-lift), producing $178$ unique \textit{IMUGait} patterns. Each pattern attacked a wide variety of user authentication models. The deeper analysis of error rates (before and after the attack) challenges the belief that authentication systems based on IMUGait patterns are the most difficult to spoof; further research is needed on adversarial models and associated countermeasures.
\end{abstract}





\maketitle

\section{Introduction}
Gait or walking patterns have been extensively researched for user recognition. Various methods can be employed to capture gait patterns, including vision-based approaches (2D/3D/infrared cameras) \cite{VideoGaitInfraredGait}, floor sensors such as pressure sensors \cite{FloorSensorGait1, FloorSensorGait2}, Inertial Measurement Unit (IMU) consisting of accelerometers, gyroscopes, and magnetometers \cite{PhoneGait1, PhoneGait2, CovariatesOrientation1}, acoustic sensors \cite{AudioGait1, AcousticID}, and Wi-Fi signals \cite{WiFiGait} among others. Some of them, e.g., video/infrared cameras and floor sensor-based gait patterns, are suitable for surveillance in controlled environments such as military bases and airports because minimal user cooperation is required. In contrast, gait patterns captured via IMUs necessitate user cooperation, making them appropriate for securing access to private properties such as cars, doors, phones, and lockers.

Early research involving IMUs in user recognition include Bamberg et al. \cite{TheFirstAccBasedGait0}, Mantyjarvi et al. \cite{TheFirstAccBasedGait0, TheFirstAccBasedGait1}, Gafurov et al. \cite{TheFirstAccBasedGait2, 2007ZeroEffortGafurov}, and Rong et al. \cite{PortableDeviceGait2}, who primarily utilized accelerometer sensors embedded in chips or portable devices, achieving reasonably low classification error rates. These studies also pointed to limitations on the applicability of IMU-based user recognition to real-world environments \cite{CovariatesAffectGait2008, CovariatesSpeedSurfaceAffectGait}. Exploration of IMU-based gait patterns for user recognition increased in 2010 \cite{PhoneGait1, PhoneGait2, WearableGait2011, Hestbek2012, Derawi2013} with the immense popularity of commercial phones which contained IMUs. Researchers also explored IMUs built into smartwatches \cite{SmartWatchWisdom, ArmMovement} in 2016. They found it even more suitable and accurate than IMUs built into smartphones, primarily due to the watch sensors' fixed position (wrist) on our bodies. 

Previous studies mostly reported results on the data collected in controlled environments such as lab settings. Changing contexts and covariates increased intra- and inter-user variances negatively affect recognition performance. Device placement, speed, orientation, walking surfaces, footwear, and backpacks are major obstacles to commercializing IMU-based gait technology \cite{RossSurvey2018}. Over the past decade, researchers have effectively addressed or investigated many of these issues \cite{InertialSensorSurvey2015, RossSurvey2018, CovariatesPace1, CovariatesPace22012, CovariatesOrientation1, CovariatesOrientation2, CovariatesOrientation3, CovariatesOrientation2014Invariant, ContextAware2014}. For instance, pace-related concerns were tackled in \cite{CovariatesPace1, CovariatesPace22012}, orientation-related issues were addressed in \cite{CovariatesOrientation1, CovariatesOrientation2, CovariatesOrientation3, CovariatesOrientation2014Invariant}, and context-awareness (phone-in-pocket and phone-in-hand) was resolved using a multi-template framework by Primo et al. \cite{ContextAware2014}. These studies have only strengthened the argument on the feasibility of deploying IMU-based gait recognition technology. 

The security of IMU-based gait recognition from active adversaries needs more attention. Like other biometric systems, the IMU-based gait recognition pipeline consists of five components: sensor, feature extractor, template database, pattern matcher, and decision-maker (ISO /IEC JTC1 SC37 SD11) \cite{AttackOnBiometricSystems2, ISOBiometricStd, JainBiometric}. Circumvention of such systems is possible by faking (imitating or mimicking) the biometric \cite{SerwaddaRobotic}; replaying stolen biometric samples \cite{TypingAttack1}; and overriding the feature extractor, the extracted features, pattern matcher, the match score(s), decision-maker, threshold, or the decision \cite{AttackOnBiometricSystems2, TypingAttack1, AttackSources}. 

There exist a variety of techniques to prevent some of these circumvention techniques. For example, overriding-based circumvention techniques can be prevented by adopting encrypted communication channels and biometric protection techniques \cite{BiometricProtectionTechniques, ISOBiometricStd}. Also, it is difficult to execute the override and replay-based attacks as they usually require a write-permission to the authentication component(s) \cite{SerwaddaRobotic}. Once an adversary gets a write-permission to the device design, an attack would be pointless \cite{PhoneSwiping1, SwipingAttack3}. On the contrary, circumvention by imitation does not require a write-permission to the device and is immune to template protection techniques and encrypted communication channels \cite{AttackSources, TypingAttack3, TypingAttack1}. Circumvention by imitation might require either stolen biometric samples from the targeted user \cite{SerwaddaRobotic} or a huge database of biometric samples \cite{TypingAttack3}. Neither is it impossible to obtain biometric samples from a targeted user nor download a publicly available database \cite{TypingAttack3, SwipingAttack3}. Most imitation-based circumvention studies reported that IMU-based gait is one of the most difficult traits to imitate \cite{2006FirstAttackGafurov, 2007GafurovAttackGender, GafurovAttack2007, GafurovAttack2009, StangAttack2007, 2010MjaalandAttack, MjaalandAttack2010Plateau, Mjaaland2009GaitMAThesis, MuaazAttack2017, PredictabilityAndResilienceOfGait}. The reason cited by these studies revolved around the inability to teach imitators (even trained actors) to copy someone else's walking pattern due to the imitator's own physiological and psychological boundaries. Purposeful and repeated circumvention of \textit{IMUGait} required for defeating continuous recognition of users is even more difficult \cite{RossSurvey2018}.

However, studies such as \cite{StangAttack2007, kumar2021treadmill, TreadmillAttack, mo2022ictganan} have suggested otherwise, i.e., it is possible to circumvent IMU-based gait recognition if adversaries have access to the resources, they put in the required effort, and if there is strong motivation to do so. Considering the National Institute of Standards and Technology guideline on \textit{effort} as a factor in circumvention attempts \cite{EffortAsAFactorInBiometricAttack}, the circumvention of \textit{IMUGait} can be divided into three categories viz. \textit{zero-effort}, \textit{minimal-effort}, and \textit{high-effort}. The basis for categorization is the number of attempts or amount of effort imitators make and the level of training and assistance they receive during the imitation process. The terms \textit{zero-effort} (or friendly scenario, or random attempt) and \textit{minimal-effort} (or hostile scenario) have been used in the past studies \cite{2006FirstAttackGafurov, GafurovAttack2007, 2007GafurovAttackGender, GafurovAttack2009}. The \textit{zero-effort} imitation refers to the scenarios in which the imitator makes no deliberate attempt to imitate the targeted user. In contrast, the \textit{minimal-effort} imitation attempt is about choosing the imitators carefully (e.g., trained actors or people of similar physical characteristics or gender as of the target) and deliberately attempting to copy the targeted user's IMU-based gait patterns. The amount of training and the number of imitation attempts in the minimal effort could be limited to walking side by side, watching videos or plots of accelerometer readings, and making not more than a couple of imitation attempts. On the other hand, \textit{high-effort} mimicry may refer to the cases in which the imitators are trained for days to weeks via statistical feedback in addition to streaming videos (or verbal feedback) \cite{MjaalandAttack2010Plateau, 2010MjaalandAttack, Mjaaland2009GaitMAThesis, TreadmillAttack}. The \textit{high-effort} mimicry might also refer to designing a robot or assisting human imitators with machines like a treadmill in addition to video-based (or verbal) feedback \cite{kumar2021treadmill, TreadmillAttack}.  

The high-effort attacks \cite{kumar2021treadmill, TreadmillAttack} mainly utilize a feedback loop-based mechanism that requires the attackers to train an individual to mimic another individual. As the name suggests, launching a high-effort attack could be tedious in practice compared to minimal and zero-effort attacks. Nonetheless, designing high-effort attacks has provided insight that this paper exploits to develop a comparatively more practical attack than the feedback loop-based training process. The insight was that individuals' walking patterns at different controllable and adaptable factors that dictate the IMU-based gait patterns form a gait spectrum. The "gait spectrum" consists of various reproducible gait patterns and can be seen as a dictionary of IMU-based gait passwords. We hypothesized that the dictionary of IMU-based gait patterns, if it includes sufficiently enough factors coupled with physical and demographically diverse groups of individuals, could help (1) systematically find imitators who will most likely be able to imitate--making the feedback loop-based imitation process much easier and (2) reproduce the samples that are needed to fool even the continuous verification system, at attacker's will. 

To this end, we summarize the key contribution of this paper as follows: 

\begin{itemize}
    \item We created a dictionary of $178$ unique \textit{IMUGait} patterns, each containing more than 100 steps of walking data, by recruiting nine physically and demographically diverse individuals who walked at varying levels (degrees) of four gait factors.
    \item We created $75$ baseline models for each of the $55$ genuine users utilizing five classifiers and all possible combinations of four sensors available in the IMUs of the smartphone
    \item We evaluated the impact of the proposed dictionary attack on these baselines at classifier and used levels. 
\end{itemize} 

The rest of the paper is organized as follows. Section \ref{RelatedWorks} describes the closely related works and distinguishes them from the presented work, Section \ref{ExperimentalDetails} lays out the experimental details, Section \ref{ResultsDiscussion} discusses the results and limitations, and Section \ref{Conclusion} concludes the paper with possible future directions for extension \footnote{Codebase: https://github.com/rajeshjnu2006/DictionaryAttackOnIMUGait}. 


\section{Related work}
\label{RelatedWorks}
Prior studies on \textit{zero-} and \textit{minimal-} effort attacks are \cite{2006FirstAttackGafurov, GafurovAttack2007, 2007GafurovAttackGender, GafurovAttack2009, MuaazAttack2017} while those that fall under the \textit{high-effort} category are \cite{StangAttack2007, 2010MjaalandAttack, TreadmillAttack, kumar2021treadmill}. Gafurov et al. \cite{2006FirstAttackGafurov, GafurovAttack2007, 2007GafurovAttackGender, GafurovAttack2009} conducted a series of experiments to evaluate the security of \textit{IMUGait}-based authentication systems against imitation. The study sequence suggested that imitation is not a substantial threat; however, individuals of the same gender or the closest person in the database could be a potential threat. For a wider coverage of zero- and minimal-effort attacks, we refer to the related work section of Kumar et al. \cite{kumar2021treadmill}, which covers \cite{2006FirstAttackGafurov, GafurovAttack2007, 2007GafurovAttackGender, GafurovAttack2009, 2010MjaalandAttack, MuaazAttack2017} in sufficient detail. Here, we summarize previous high-effort attack attempts \cite{StangAttack2007, 2010MjaalandAttack, Mjaaland2009GaitMAThesis, MjaalandAttack2010Plateau, TreadmillAttack, MuaazAttack2017, kumar2021treadmill}. 

The first experiment under the \textit{high-effort} category was conducted by Stang et al. \cite{StangAttack2007} in 2007. Stang et al. \cite{StangAttack2007} provided plots of accelerometer readings in $x$, $y$, and $z$ dimensions on a big screen and a match score (Pearson's correlation coefficient computed between the resultant acceleration of the imitator and the target) as feedback. The study analyzed five gait templates from one user, collected under different speed and step length settings. Thirteen imitators attempted to match each template fifteen times without prior knowledge of the target's walking patterns. Some imitators surpassed a 50\% match, suggesting that gait patterns could be imitated with rigorous training. However, Mjaaland et al. \cite{2010MjaalandAttack, Mjaaland2009GaitMAThesis, MjaalandAttack2010Plateau} criticized the study's conclusion due to its limited sample size (only one individual) and the reliance on Pearson's correlation coefficient to measure success.

Mjaaland et al. \cite{2010MjaalandAttack, Mjaaland2009GaitMAThesis, MjaalandAttack2010Plateau} conducted a study to determine whether extensive training could improve an imitator's ability to mimic a target's gait. They used regression analysis to analyze the imitators' learning curves and divided the imitation process into three scenarios: friendly, short-term hostile, and long-term hostile. The friendly scenario served as a baseline and involved collecting regular walking patterns from 50 participants, resulting in 6.2\% of Equal Error Rate (EER)-- a point on the Receiver Operating Characteristics point where both false acceptance rate and false rejection rate are the same. In the short-term hostile scenario, one target and six imitators were chosen based on gait stability, Dynamic Time Warping (DTW) distance, and eagerness to participate. Each imitator made five attempts, with feedback provided between sessions. Despite the training, no imitator could circumvent the authentication system, as they could not breach their physiological boundaries and match all the target's traits. Over-training led to more unnatural and mechanical walking patterns. In the long-term hostile scenario, one imitator was selected for six weeks of training. While the imitator showed multiple plateaus, the uncertainty and insufficient data prevented the authors from having any strong conclusions. The study highlighted that statistical feedback was more helpful than visual feedback and that simultaneously concentrating on different gait factors was challenging. The authors concluded that gait imitation is a difficult task.

Muaaz et al. \cite{MuaazAttack2017} conducted a study in 2017 to investigate the resilience of \textit{IMUGait} authentication systems to impersonation attempts. They examined three circumvention scenarios: zero-effort, reenact, and coincide. The zero-effort scenario collected data from 35 participants, with user-specific models trained, considering the rest as impostors. Under the reenact and coincide scenarios, nine of the 35 participants participated in the impersonation process, with five acting as imitators and four as victims. The imitators were trained mime artists skilled in mimicking body motions and language. In the reenact phase, imitators observed and rehearsed the target's gait movements for 10 minutes. In the coincide phase, they received live feedback comparing their gait to the target's. Using Dynamic Time Warping (DTW) distances to measure success; no attacker achieved the threshold that genuine users easily attained. The best attempt matched only 25\%, well below the 50\% threshold. The system removed 27\% of imitation attempts as outliers, concluding that the more imitators tried, the more outlying gait cycles they produced. The results suggested that \textit{IMUGait} authentication is resilient to impersonation-based circumvention attempts.

Kumar et al. \cite{TreadmillAttack, kumar2021treadmill} focused on producing the sensor readings rather than reproducing a visually similar walk to the target. They observed critical gait factors that can be controlled and adapted and strongly correlate with the features extracted from IMUs readings. They used a treadmill to breach the physiological boundaries of the imitators and developed a feedback-based attack model, a proven mechanism to alter human behavior, to train the imitators. They reported that the attack increased the average false accept rates from 4 to 26\% on a dataset of 18 users with just one imitator.  

Building on the previous studies and motivated by dictionary attacks on PIN and password-based authentication systems, our work emphasized creating a dictionary of IMU-based gait patterns. The patterns in the dictionary were produced by nine carefully recruited individuals who walked at different speeds, step lengths, step widths, and thigh lifts, producing $178$ unique gait patterns, each with a minimum of $100$ steps. The attack with the closest sample in the dictionary increased the average false accept rates to 32-40\%. Such an increase from 6-14\% to 32-40\% raises a serious question about the security offered by IMU-based gait authentication.  

\section{The proposed attack}
We present a novel threat model to evaluate the inherent vulnerabilities of IMU-based gait authentication systems. There are two actors in the attack process: the \textit{target} users and \textit{attackers}. The \textit{target} users are legitimate users whose biometric gait data is used to train machine learning classifiers for authentication. \textit{Attackers}, on the other hand, are malicious entities attempting to gain unauthorized access by mimicking the gait patterns of legitimate users. We assume the attacker has a \text{dictionary} of pre-recorded gait patterns. The dictionary's keys consist of the details about the imitator and the settings at which the corresponding values were produced. We assume that given a \text{target} gait pattern, the \text{attackers} would be able to find the closest gait pattern in the dictionary based on a predefined metric and ask the person associated with that gait pattern to reproduce the pattern in real-time by walking at the saved settings, to deceive the system. Our findings (refer to Section \ref{uservsuniqueattacks_a_svm}) show that more than one \textit{attackers} could trigger false acceptance with real-time physical behavior. We assume the following capabilities for the attacker:

\textit{Knowledge of System Architecture} The attacker need not know the machine learning algorithms and parameters used in the authentication system.

\textit{Data Collection} The attacker can acquire a range of gait patterns either by breaching databases or by other illicit means, such as luring users to install an app on their phone for a discount or something similar.

\textit{Device Access} The attacker will have physical access to the device for the attack period.

The proposed end-to-end attack identifies theoretical vulnerabilities and empirically validates them by simulating real-world attack scenarios. The empirical findings indicate that attackers with these capabilities pose a significant risk to the security of IMU-based gait authentication systems, necessitating the development of countermeasures.

\section{Design of experiments}
\label{ExperimentalDetails}
\subsection{Genuine dataset}  
\label{SubsecGenuineDataset}
\begin{figure}[htp]
    \centering
    \includegraphics[width=3in, height=2in]{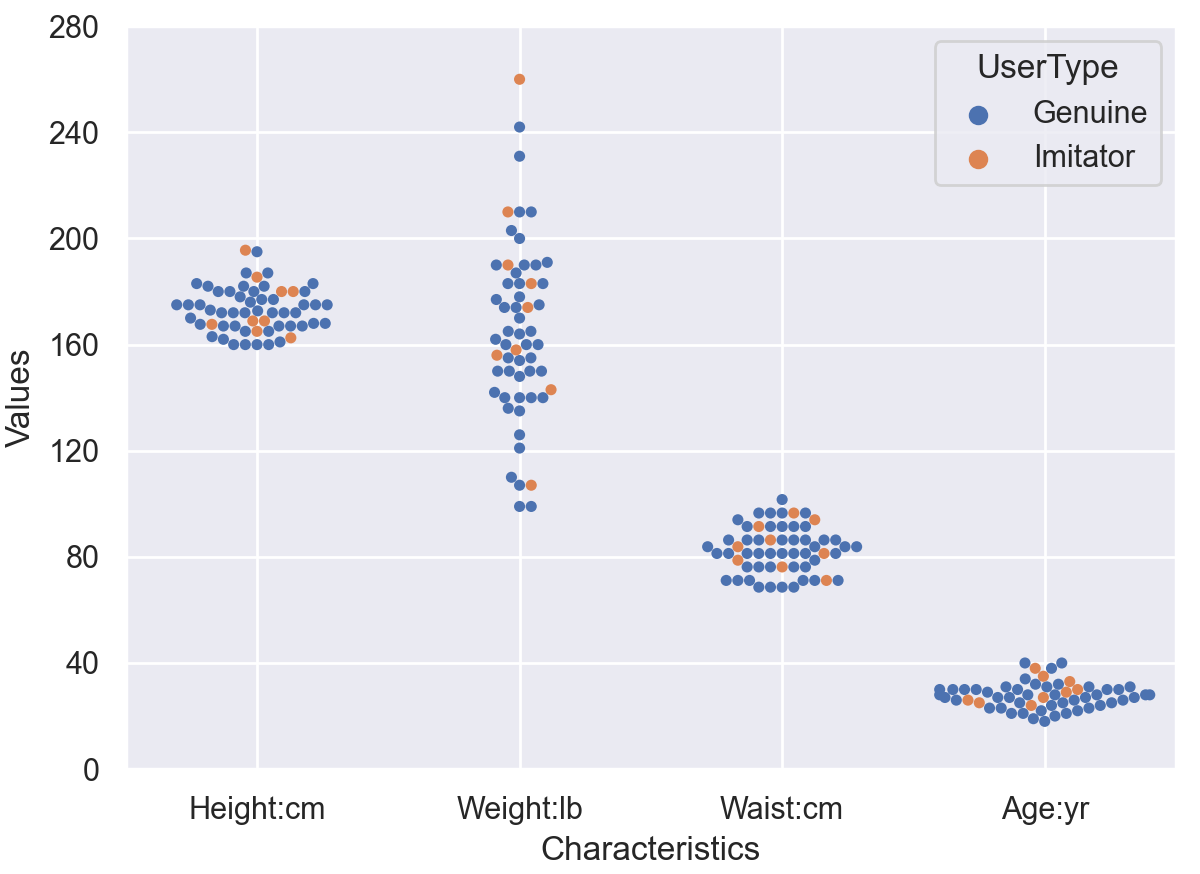}
    \caption{Physical characteristics of genuine users and selected imitators. We could gather the physical characteristics of only $48$ genuine users. Thus, there are only $48$ blue dots, while nine orange dots indicating impostors are plotted for each characteristic.}
    \label{GenuinImitatorDemoGraphicCom}
\end{figure}

The Genuine dataset consists of data collected from individuals who walked naturally in a $328$ feet long and $6.5$ feet wide corridor back and forth. The demographic and physical attributes of the participating individuals who ($48$ of $55$) provided their details are in Figure \ref{GenuinImitatorDemoGraphicCom}. The data was collected through an Android app installed on an Android phone. The phone was always placed in the right pocket of the participant's trousers. The participants freely (without any instruction) walked two to three rounds in the corridor. The participants then repeated the same exercise after a few minutes, hours, or days later, depending on their availability. The idea behind collecting data in separate sessions was to keep the training and testing data separate. 

The data collection App was designed to collect data via four specific sensors, namely, \textit{linear\_acceleration}, \textit{gyroscope}, \textit{magnetic\_field}, and \textit{rotation\_vector} of the Android platform. The \textit{linear\_acceleration} was used to record the phone's acceleration in three physical dimensions (x, y, and z), excluding the force of gravity. The \textit{gyroscope}  captured the angular rotation of the device in radian/second around three physical axes (x, y, and z) of the device. The \textit{magnetic\_field} was used to record the ambient geomagnetic field for all three physical axes (x, y, z) in Tesla ($\mu$). \textit{Rotation\_vector} recorded the device's orientation as a combination of an angle and an axis, in which the device had rotated through an angle $\theta$ around an axis $(x, y, z)$. The three elements of the rotation vector are $(x \times sin( \theta /2), y \times sin(\theta/2), z \times sin(\theta/2))$, such that the magnitude of the rotation vector is equal to $sin(\theta/2)$. The direction of the rotation vector is equal to the direction of the axis of rotation. In this case, the reference coordinates include East, North Pole, and Sky directions. 

Overall, the Genuine dataset consists of gait patterns collected from fifty-five individuals. Each individual provided around $320$ gait cycles in two separate sessions. As a result, the Genuine dataset consisted of $17,600$ (=$55 \times 320$) gait cycles in total. Since the authentication models were implemented using multi-class classification algorithms, they required samples (or feature vectors) from genuine and impostor classes. Following previous studies \cite{TrainingTestingUsingOtherUsers, PhoneGait1, PhoneGait2, CovariatesOrientation1, PhoneGait7, PhoneGait6, ContextAware2014, PhoneMovementPrinceton, SandeepADeauth}, we used samples from users other than the genuine as impostor samples for evaluating the performance of the baseline authentication models (i.e., under \textit{zero-effort} attacks).

\subsection{Dictionary dataset}
\label{DictionaryDataset}
\begin{figure}[htp]
    \centering
    \includegraphics[width=3in, height=1.85in]{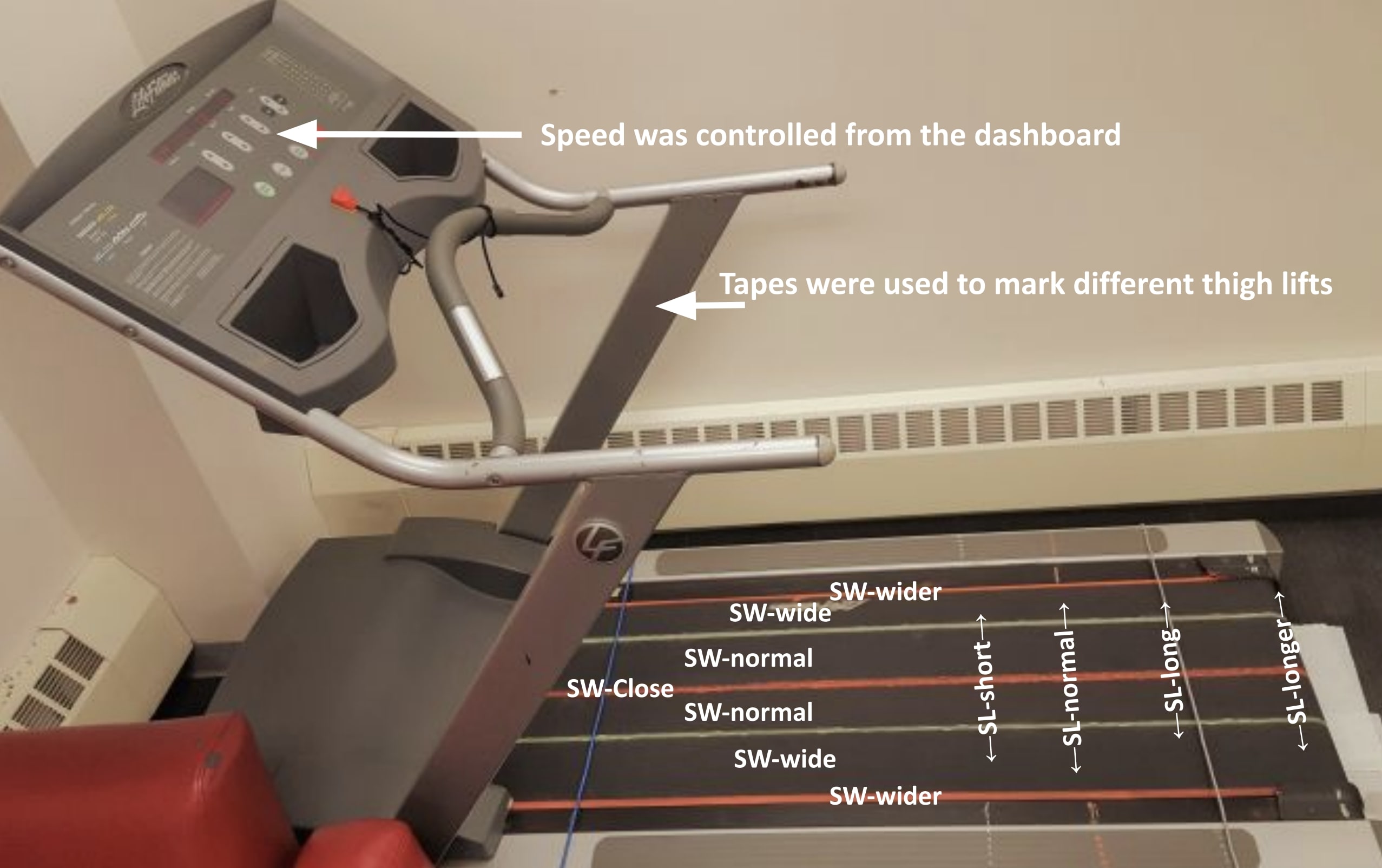}
    \caption{Dictionary data collection setup. Chalk and wire markers show different step lengths and width settings. The speed controller is on the dashboard as indicated, and the thigh lift markers are on the limbs of the treadmill (invisible in this view of the picture).}
    \label{DictionarySetup}
\end{figure}

For creating the \textit{Dictionary dataset}, we recruited nine imitators following IRB approval from the university. The imitators were carefully chosen, keeping their physical characteristics (height, weight, age, waist, and gender) in mind. The aim was to create a spectrum of IMU-based gait patterns as wide as possible. Any pattern on the spectrum can be reproduced on demand. The requirement of the on-demand reproducibility of these patterns motivated us to store the data in the form of a dictionary in which the keys represented the tuple of the imitator id and a set of four values representing the levels of the four gait factors (speed, step length, step-width, and thigh lift) that are easy to control, adapt, and recreate with access to a treadmill. The dictionary values, on the other hand, represented the gait patterns produced for the corresponding keys. The recruited imitators walked at possible (to the imitators) variations of four gait factors (speed, step length, step width, and thigh lift). Six imitators walked at $21$ variations, two at $16$ variations, and one at $20$ variations of the four gait factors on a treadmill totaling to $178$ $(=21\times6 + 1\times20 + 2\times16)$ unique gait patterns. The Speed (SP) varied from 1.4 miles per hour ($63$ centimeters/second) to $3.0$ miles per hour ($134$ centimeters/second) at the interval of $0.2$ miles per hour ($9$ centimeters/second), constituting a total of $9$ speed settings. The step length (SL) varied from short to normal, long, and longer. The step width (SW) varied from close, normal, wide, and wider. Similarly, thigh-lift (TL) varied from back, normal, front, and up. Each entry in the dictionary consisted of at least $100$ steps summing to at least $8,900$ $(=178 \times 50)$ gait cycles in the Dictionary-effort dataset. The experimental setup is shown in Figure \ref{DictionarySetup}. 


\begin{figure*}[htp]
  \centering
  \begin{tabular}{cc}
    \includegraphics[width=3.2in, height=1.6in]{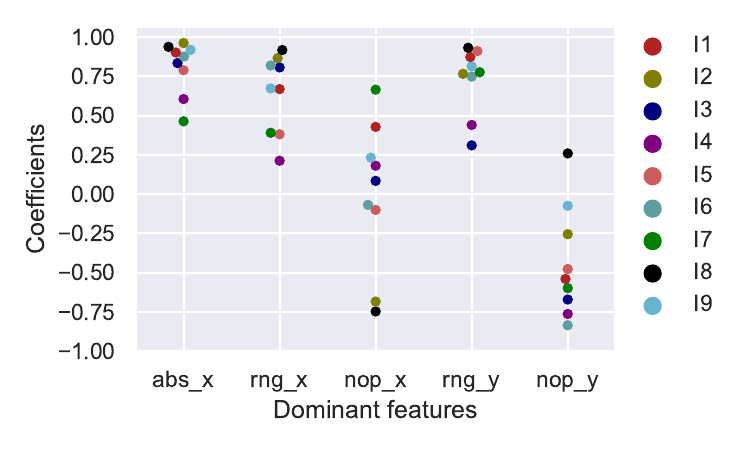}&
    \includegraphics[width=3.2in, height=1.6in]{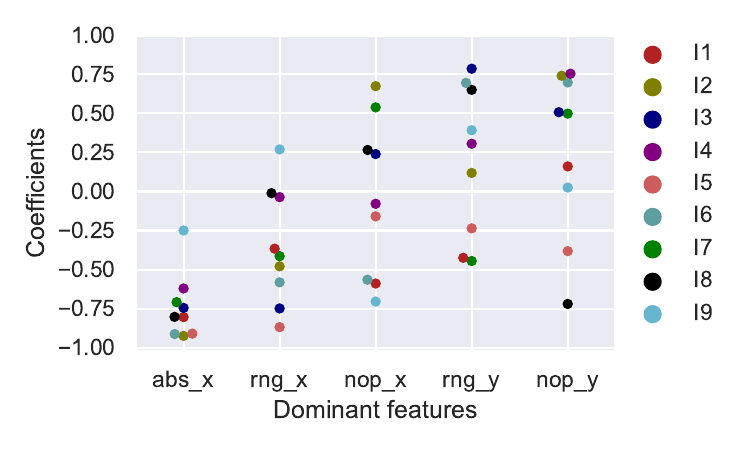}\\
    \includegraphics[width=3.2in, height=1.6in]{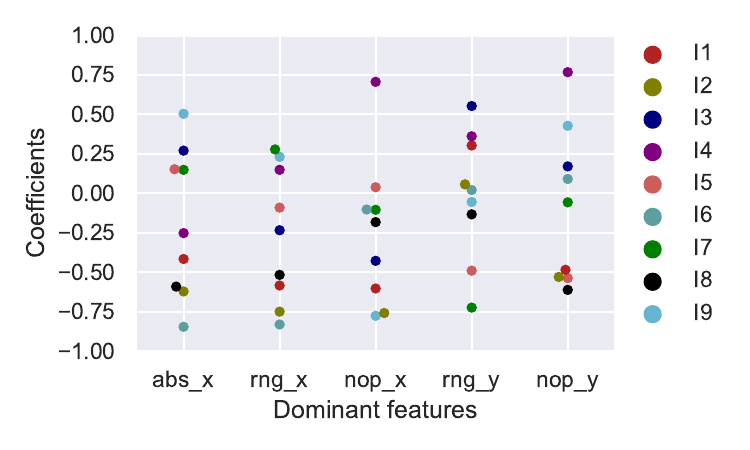} &
    \includegraphics[width=3.2in, height=1.6in]{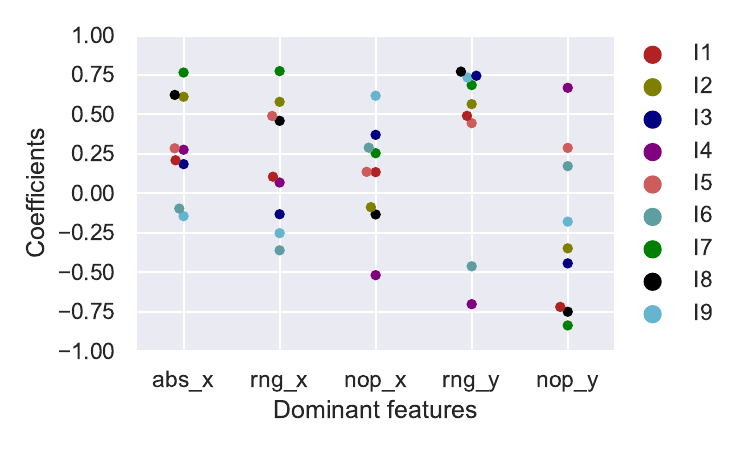}
  \end{tabular}
  \caption{Upper-left: speed, Upper-right: Step-length, Lower-left: Step-width, Lower-right: Thigh-lift. For nine imitators, EDA revealed an interesting relationship between the four gait factors and dominant features (defined by \cite{TreadmillAttack}). We can observe that adjusting the same gait factor for different imitators affects the feature values differently. This provides a pictorial insight into why \textit{IMUGait} patterns are unique for different users. For example, we can look at the upper left figure; the impact on the feature named $nop\_x$ is the opposite for a change in speed for imitators I7 (correlation coefficient is +0.7) and I8 (correlation coefficient is -0.75). Similarly, in the bottom-right figure, the impact on the feature names $nop\_y$ significantly differs for imitators I4 and I7. More than 77\% of the correlations were statistically significant at $\alpha =0.05$, indicating a strong relationship between gait factors and features.}
  \label{ImitationProfiles}
\end{figure*}

\subsection{EDA on Dictionary dataset} Exploratory data analysis (EDA) was conducted on the Dictionary dataset to find out (1) how varying the levels of gait factors would impact the dominant features defined by Kumar et al. \cite{TreadmillAttack} for different imitators, (2) whether the gait patterns collected under the same settings were significantly similar, and (3) whether the gait patterns collected at different settings differed from each other differently. For analyzing (1), we used the Pearson correlation coefficient as the feature variables were continuous, and factors were either continuous (e.g., speed) or ordinal (step length, step width, and thigh lift). Figure \ref{ImitationProfiles} demonstrates the intrinsic relationship unearthed by the correlation analysis between the gait factors and prominent features. Images in Figure \ref{ImitationProfiles} suggest why IMU-based gait biometric is so unique among individuals, as we can see that a change of the same magnitude in any gait factor can affect the feature values differently for different imitators—a motivation to recruit diverse imitators if one wants to increase the chances of success of the dictionary attack. 

To investigate (2) and (3), we used histogram intersection, defined as follows, 

$\mathbb{I}(P, Q) = \frac{\sum_{i=0}^{n} min (P_i, Q_i)}{\sum_{i=0}^{n} P_i}$, where $P$ and $Q$ are histograms, each containing $n$ bins. 

\begin{figure*}[htp]
    \centering
    \includegraphics[width=5.8in, height=2.6in]{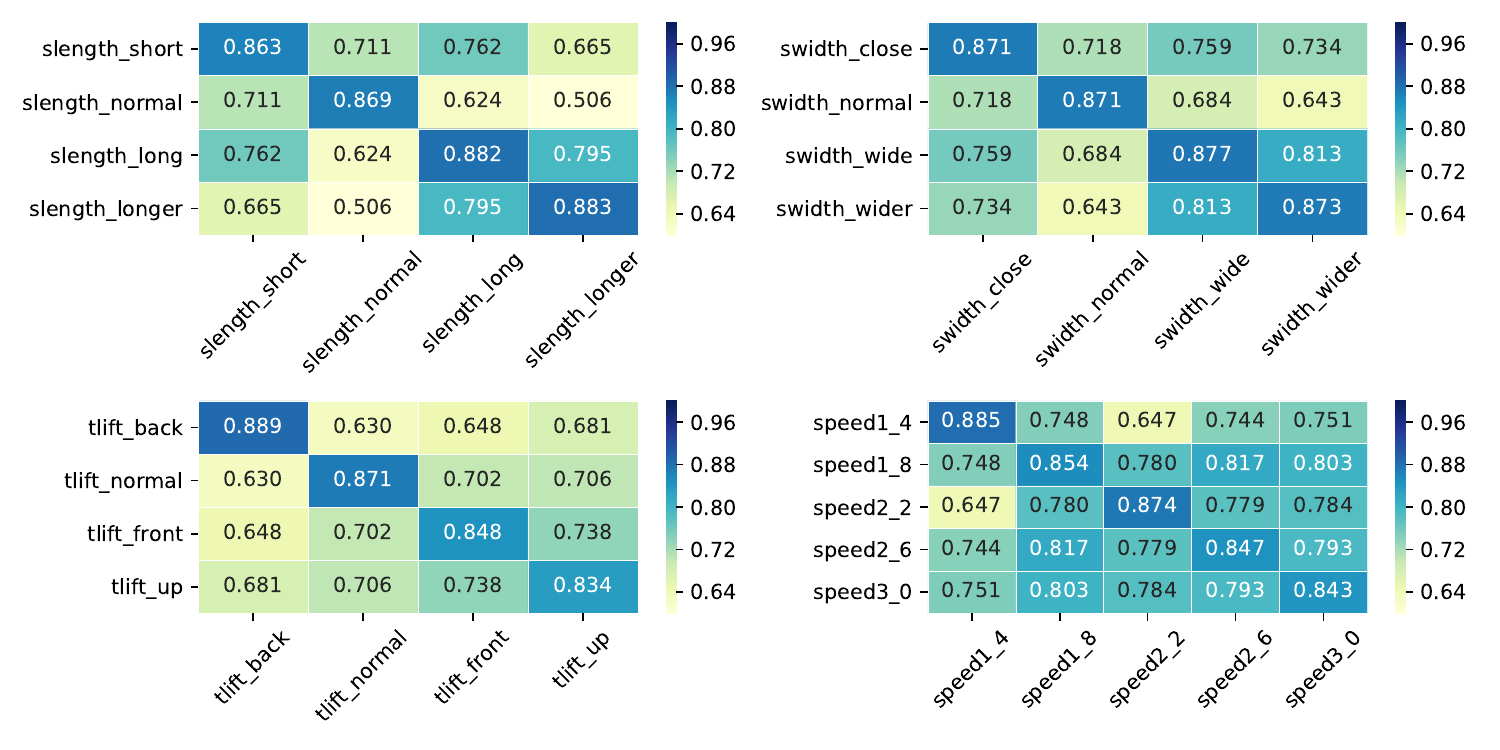}
    \caption{These heatmaps demo the overlap of the data in different frames created from the raw sensor data collected in the same and different settings. The same settings overlap was computed using windows of data collected at different times. In general, we observed higher histogram overlap for the data collected in the same settings of individual gait factors compared to the data collected in different settings of the gait factors in most cases. For example, the upper-left heatmap suggests that the window of data collected at short step length has a higher overlap (0.863) with the data collected at the same setting at a different time compared to the data collected at normal (0.711), long (0.762), and longer (0.665) step lengths. Similarly, the lower-left heatmap suggests that the window of data collected at thigh-lift back highly overlaps (0.889) with the window of data collected at the same setting but at different time intervals compared to the overlap with the windows of data collected at thigh-lift normal (0.630), front (0.648), and up (0.681). The plotted numbers are the average histogram intersections computed over at least nine windows of $8$ seconds of data for the accelerometer's x-axis. Each of the histogram computations used $80$ bins of equal width. We observed similar phenomena for different axes of the same and different sensors.}
    \label{I7_hist_intersection}
\end{figure*}

The overlap of probability distribution (histograms) of data produced at the same level was much higher ($>0.85$) compared to the data produced at different levels of gait factors for the same imitator (see Figure \ref{I7_hist_intersection}). We observed similar phenomena across the axis, sensors, and imitators with random inspection. This indicates that the \textit{IMUGait} patterns produced at different settings are different and capture unique regions on the human gait spectrum. 

\subsection{Baseline \textit{IMUGait} authentication pipeline}
\subsubsection{Data segmentation}
An authentication system is desired to have high accuracy and low decision latency. One of the ways to ensure quicker decisions is segmenting the data into smaller frames while maintaining distinctiveness. Two widely studied approaches are cycle \cite{2006FirstAttackGafurov, GafurovAttack2007, 2007GafurovAttackGender, GafurovAttack2009, 2010MjaalandAttack, MuaazAttack2017} extraction-based segmentation and sliding window-based fixed-length frame \cite{InertialSensorSurvey2015, PhoneMovementPrinceton, kumar2021treadmill} extraction. Studies that use cycle extraction-based approaches use a point-wise comparison of the train and test samples using classical distance measures such as Euclidean distance and \textit{DTW}. On the other hand, studies that use fixed-length-based frame (both overlapping and non-overlapping) extraction schemes use a variety of machine learning classifiers. Thang et al. \cite{PhoneGait7} specifically investigated both approaches and concluded that the frame plus machine learning-based approach achieved significantly better results $(92.7\%)$ than the cycle-based approaches $(79.1\%)$ on the same dataset. Besides, Al-Naffakh et al. \cite{FrameIsBetterThanCycleInGait} compared 29 studies and concluded that the frame-based approach almost always beats the cycle-based approaches. Following the recommendation of \cite{ArmMovement}, we applied the sliding window-based frame extraction approach for segmentation in our experiments. The window length and sliding interval dictate the delay in the first and subsequent authentication decisions. The smaller the window size and sliding interval, the quicker the decisions; thus, we used $8$ seconds of the window and $4$ seconds of sliding interval across the experiments. This process resulted in $43$ feature vectors for genuine data and $18$ for each of the $178$ patterns in the dictionary, on average. 

\subsubsection{Preprocessing}
Although the raw data plots for each user looked smooth, as these sensors are well-calibrated and corrected nowadays, we observed spikes in some places. Therefore, we applied the moving average technique to smooth the sensor readings further to avoid noisy feature values. The process of moving average-based smoothing of a signal is described below: 

Let \( X = [x(t_1), x(t_2), x(t_3), \ldots, x(t_n)] \) be a time-series signal recorded by a sensor at a specified sampling rate. This signal is assumed to contain noise. We aim to smooth this signal to obtain a less noisy version, denoted as \( X' \).

The smoothed signal \( X' \) is defined as \( [x'(t_1), x'(t_2), \ldots, x'(t_{n-s+1})] \), where \( n \) is the length of the original signal \( X \), and \( s \) is the smoothing window size.

The smoothed value \( x'(t_i) \) at each time \( t_i \) is computed as follows:
\[
x'(t_i) = \frac{1}{s} \sum_{j=0}^{s-1} x(t_{i+j})
\]
Here, \( s \) controls the number of data points taken together to update the \( i_{\text{th}} \) data point in \( X' \).
 
The value of \( s \) is determined based on the sampling rate of the original signal \( X \). After preliminary analysis, we have chosen:
\[
s = \lceil 0.05 \times \text{sampling\_rate} \rceil
\]
For example, with a sampling rate of 46, \( s \) would be calculated as \( \lceil 0.05 \times 46 \rceil = 3 \).

The smoothing equation is applied for \( i = 1, 2, \ldots, n-s+1 \). The remaining data points were used for the boundary conditions where fewer than \( s \) data points were available. 
 
An unreasonably high value of \( s \) can result in excessive smoothing, which may eliminate important high-frequency components in the signal. On the other hand, the lowest value of \( s \) (i.e., \( s=1 \)) will result in no smoothing, leaving the signal unchanged.

\subsubsection{Feature extraction and analysis}
\begin{figure*}[htp]
    \centering
    \includegraphics[width=7in, height=3.85in]{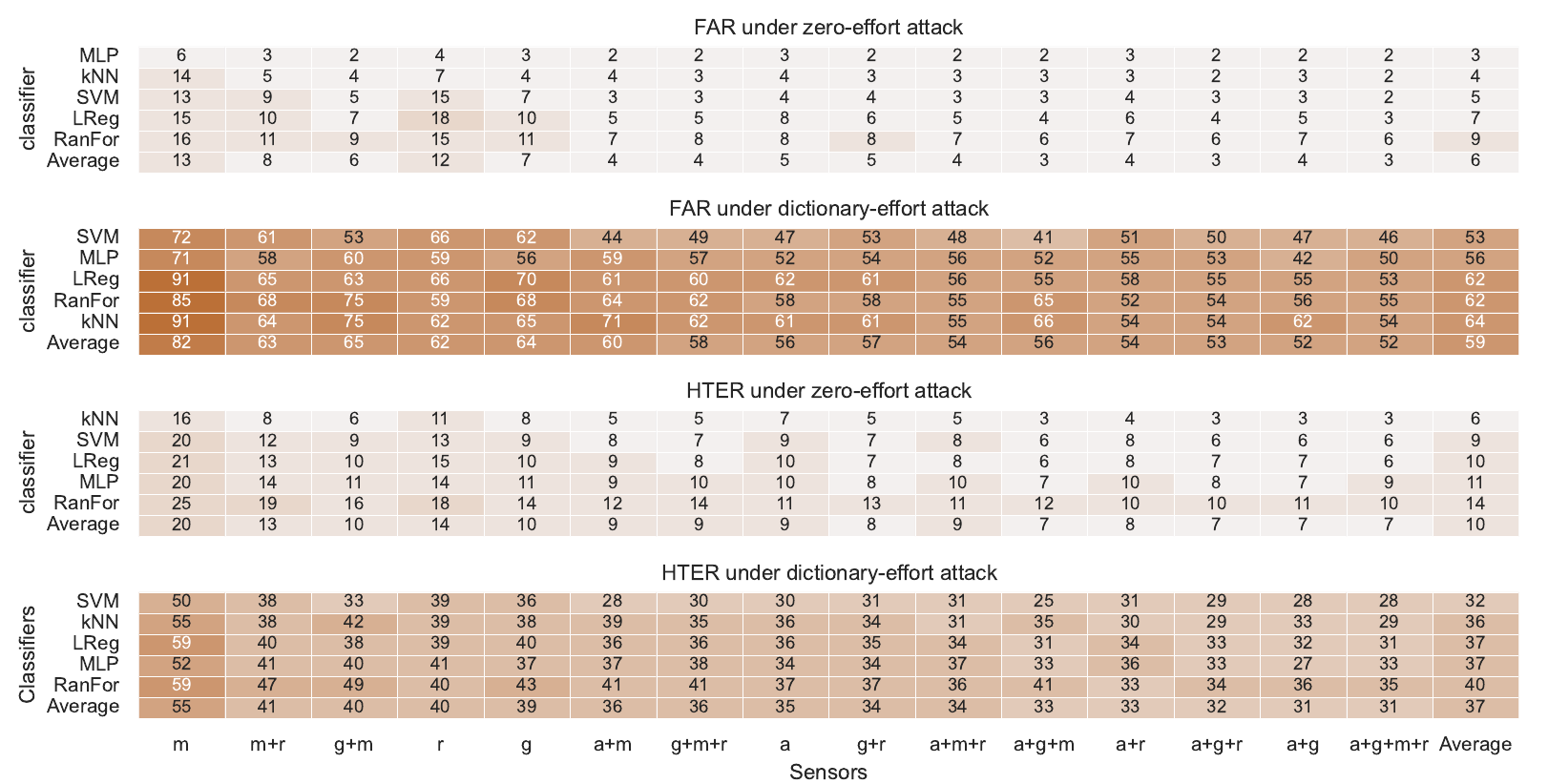}
    \caption{The mean error rates achieved by different classifiers for different sensor combinations under the zero-effort and dictionary-effort circumvention scenario. The numbers are rounded up to the nearest integer percentage. The first heatmap reports the FAR under a zero-effort attack, followed by the heatmap presenting the FAR under a dictionary-attack scenario, facilitating a glance at the damage caused. The goal of the dictionary attack is to bypass the authentication system, which means it impacts only the FAR while FRR remains as it is under the zero-effort attack. For comparison of the overall impact on the authentication systems, we report Half Total Error Rate (HTER), an average of FAR and FRR and recommended by Bengio et al. \cite{BengioWhyHTER} for both zero-effort and dictionary-effort scenarios. Notably, the presented error rates correspond to the dictionary's most successful entries (entries that caused the maximum increase in the FAR). \textit{The values are sorted on average of HTER.}}
    \label{dict-effort-sensor-level}
\end{figure*}

We extracted time and frequency domain features from each segment (frame). The time domain feature set included arithmetic mean, standard deviation, mean absolute change, mean absolute deviation, skewness, kurtosis, mean energy, the number of mean crossings, the number of peaks, first, second, and third quantiles, length of the longest strike below and above the mean, and bin counts in $16$ equally thick bins. The bin counts were inspired by \cite{PhoneGait1, SmartWatchWisdom}, and it provides different (structural) information compared to aggregate features. In comparison, the frequency-domain features included the first, second, and third quantiles and standard deviations of the Fourier transform's amplitudes. These features have been extensively studied over the years for authentication, are relatively established in the field, and have achieved significantly low error rates \cite{PhoneGait1, SmartWatchWisdom, kumar2021treadmill, PhohaGaitSurvey, GaitWearableSurvey}. This is evident via the performance of our baselines as well (see Figure \ref{dict-effort-sensor-level}).

Each raw data segment (frame) was translated into a vector of $34$ feature values, resulting in $136$ features for each sensor as we used the magnitude (computed as $\sqrt{x^2+y^2+z^2}$) beside reading on each of the axes ($x$, $y$, and $z$). Mutual Information (MI) between individual features and class labels (genuine and impostor) was computed for selecting the top $30$ features (the ones with the highest MI values). The chosen features varied across users but not significantly, as we aimed to train user-specific authentication models in which, for training each authentication model, we used genuine users' data as genuine samples while the rest of the users' data as impostor samples.

\subsubsection{Sensor fusion and classifiers} Previous studies such as \cite{kumar2021treadmill} have suggested combining different sensors helps reduce error rates and adversarial attacks' impact. We wanted to test what combination of sensors achieves the best error rates while defending itself from the proposed attack the most. Therefore, we studied all $15$ possible combinations of four sensors and reported the results. The combination was conducted at the feature level by concatenating the feature vectors before training and testing the model. The feature vectors from each sensor consisted only of the selected features from individual sensors. 
Various classification algorithms have been tested on the \textit{IMUGait} pattern for authentication. Following previous studies, we experimented with five widely applied and successful classifiers (on \textit{IMUGait} datasets): k-nearest Neighbors (kNN), Support Vector Machine (SVM), Logistic Regression (LReg), Multilayer Perceptrons (MLP), and Random Forest (RanFor), each with distinct learning paradigms \cite{GaitWearableSurvey, PhohaGaitSurvey, BodyTaps2018, OneClassISABA2018, mo2022ictganan}. 

\begin{figure*}[htp]
  \centering
  \begin{tabular}{c}
    \includegraphics[width=7in, height=2.85in]{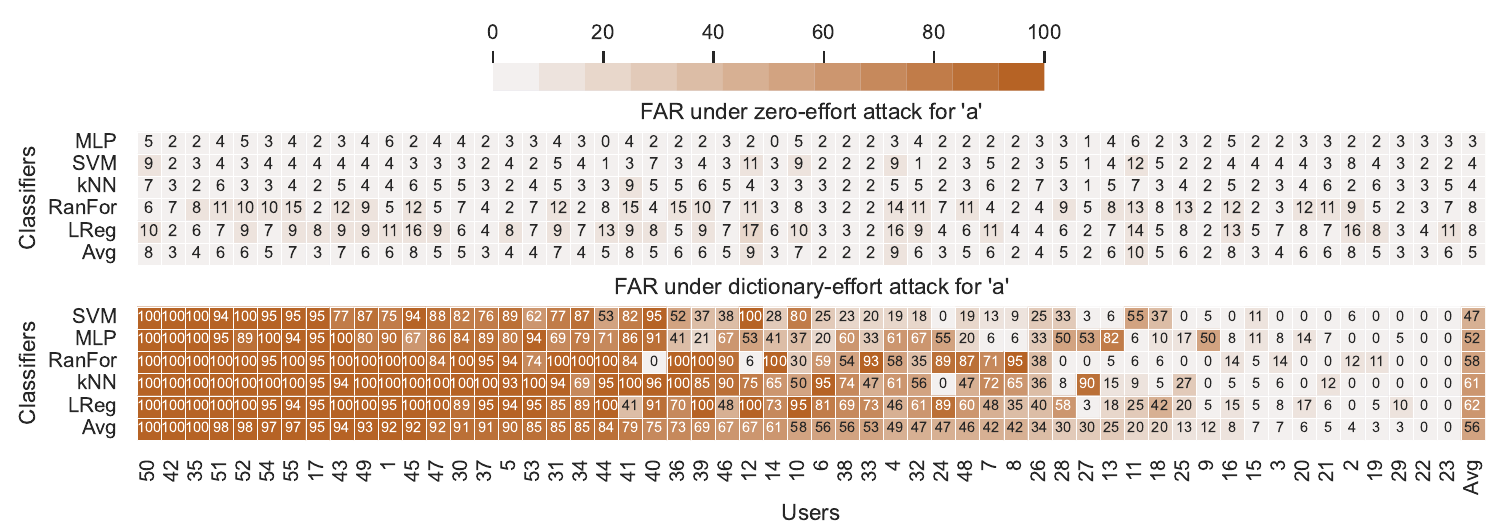}
  \end{tabular}
  \caption{The impact of Dictionary-based circumvention attempts on accelerometer-based models implemented using SVM, the most resilient classifier. The FARs and HTERs obtained under Zero- and Dictionary-effort circumvention attempts are presented one after the other for a quick comparison. These results are based on the most successful (the one that achieved the maximum increase in the FAR from the baseline) gait pattern from the dictionary. \textit{Matrix is sorted on average of dictionary FAR.}}
  \label{dict-effort-cls-user-a}
\end{figure*}

\begin{figure*}[htp]
  \centering
  \begin{tabular}{c}
    \includegraphics[width=7in, height=2.85in]{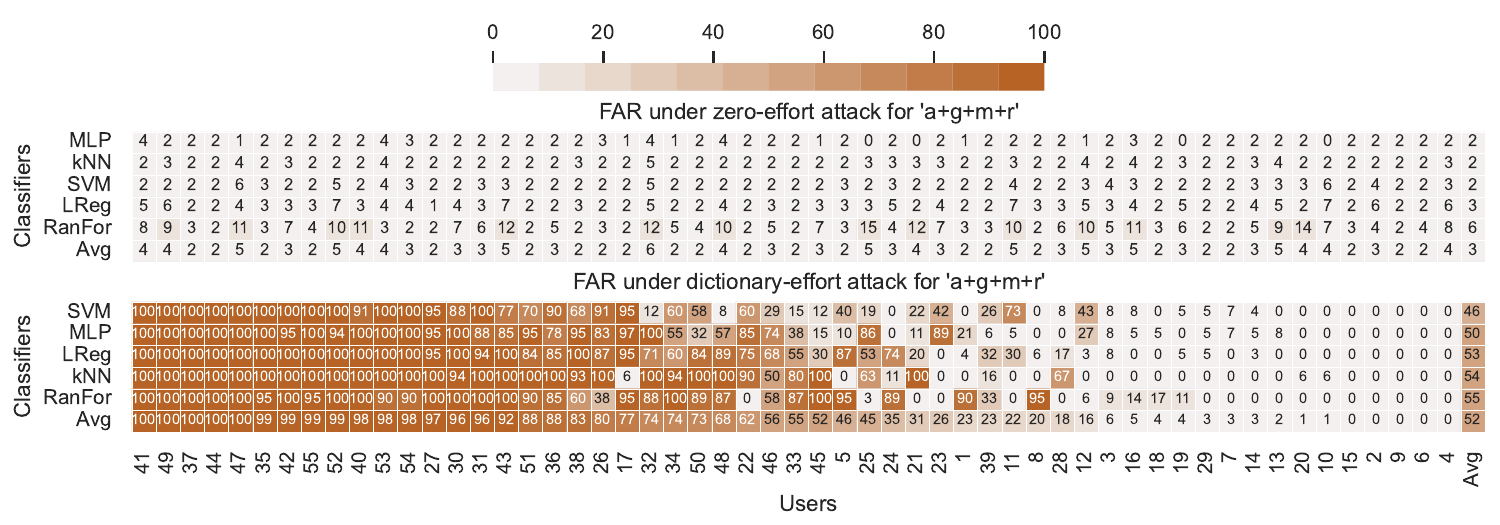}
  \end{tabular}
  \caption{The impact of Dictionary-based circumvention attempts on fusion (a+g+m+r)-based models implemented using SVM, the most resilient classifier. The FARs obtained under Zero- and Dictionary-effort circumvention attempts are presented one after the other for a quick comparison. These results are based on the most successful (the one that achieved the maximum increase in the FAR from the baseline) gait pattern from the dictionary. \textit{Matrix is sorted on average of dictionary FAR.}}
    \label{dict-effort-cls-user-all}
\end{figure*}

\subsubsection{Training, Testing, and Evaluation}
The train and test setup of the classifiers is detailed as follows. We trained the authentication model for each user separately. The Genuine dataset contained data collected in two separate sessions for every user. The first session was used for training, and the second was used for testing the authentication models. 

Formally, let $\mathbb{U} = \{u_1, u_2, ..., u_{N}\}$ be a set of $N$ genuine users in the database. The authentication model for user $u_i$ can be represented as a function $f(X_i) \mapsto \{gen, imp\}$, with $X_i$ being the feature matrix consisting of both genuine (created from $u_i$ data and impostor (created from $\mathbb{U}-u_i$) feature vectors and $\{gen, imp\}$ representing two possible classes. 

Each authentication model was tested for genuine fail rate using the genuine data from the second session and assessed in terms of False Reject Rate (FRR). Likewise, the impostor pass rate was tested using the second session data from $\mathbb{U}-u_i$ and assessed in terms of False Accept Rate (FAR). Additionally, to compare different architectures, we used Half Total Error Rate (HTER), an average of FAR and FRR, recommended by Bengio et al. \cite{BengioWhyHTER} to report the performance on the test dataset. 

\subsection{Class imbalance} While training the binary classifier-based authentication models $f(X) \mapsto \{gen, imp\}$, we used five randomly selected feature vectors from each possible impostors (i.e., $|\mathbb{U}-u_i| = 54$). As a result, we had $270$ $(=54$ impostors $\times 5)$ impostor feature vectors against $22$ (on average, per user) genuine feature vectors for training each user authentication model. To overcome the class imbalance, we over-sampled the genuine feature vector using Synthetic Minority Oversampling Technique (SMOTE) \cite{SMOTE} to match the number of impostor feature vectors. Consequently, each user authentication model has trained on a feature matrix of $540$ feature vectors. 

\subsection{Dictionary attack process}
To circumvent the trained authentication model for $u_i$, i.e., $f(X_i) \mapsto \{gen, imp\}$, one need to produce $X_{adv}$ i.e. a set of gait patterns such that $f(X_{adv}) \mapsto \{gen\}$. Previous studies \cite{GafurovAttack2007, GafurovAttack2009, 2007GafurovAttackGender, MjaalandAttack2010Plateau, MuaazAttack2017} have shown that it is not easy to generate $X_{adv}$ except for \cite{StangAttack2007, TreadmillAttack, kumar2021treadmill}. The methods proposed in \cite{StangAttack2007, TreadmillAttack, kumar2021treadmill} are tedious. This work demonstrates that $X_{adv}$ can be produced comparatively easily if we build a dictionary of \textit{IMUGait} patterns. If the entries in the dictionary are more diverse, then it is easier to find $X_{adv}$ with higher chances of success and reproducibility. 

To test the effectiveness of the proposed dictionary, we took an exhaustive approach, as we tried every entry (\textit{IMUGait} pattern) available in the dictionary to attack $f(X_i) \mapsto \{gen, imp\}$ and assessed its robustness. In a more practical scenario, the attacker would have a way of estimating the likelihood of the success of each of the entries in the dictionary and simulate the setup that produced the entry with the highest likelihood of bypassing the system.

\begin{figure*}[htp]
    \centering
    \includegraphics[width=6.9in, height=2in]{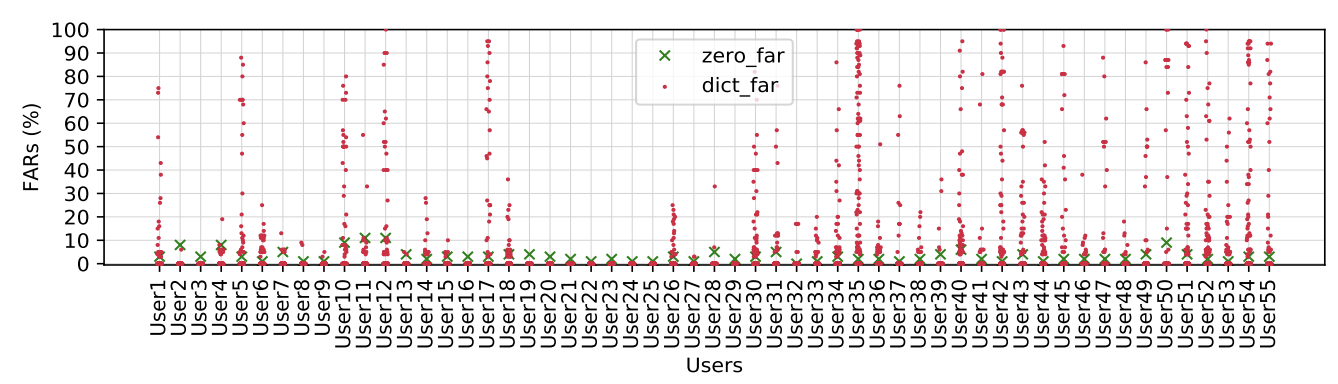}
    \caption{The distribution of \textit{FAR}s under Zero-effort attack (\textit{zero\_fars}) and \textit{FAR}s under Dictionary-effort attack (\textit{dict\_fars}) across genuine user population for accelerometer-based models respectively. $178$ \textit{IMUGait} patterns were attempted for each genuine user.}
    \label{uservsuniqueattacks_a_svm}
\end{figure*}

\begin{figure*}[htp]
    \centering
    \includegraphics[width=6.9in, height=2in]{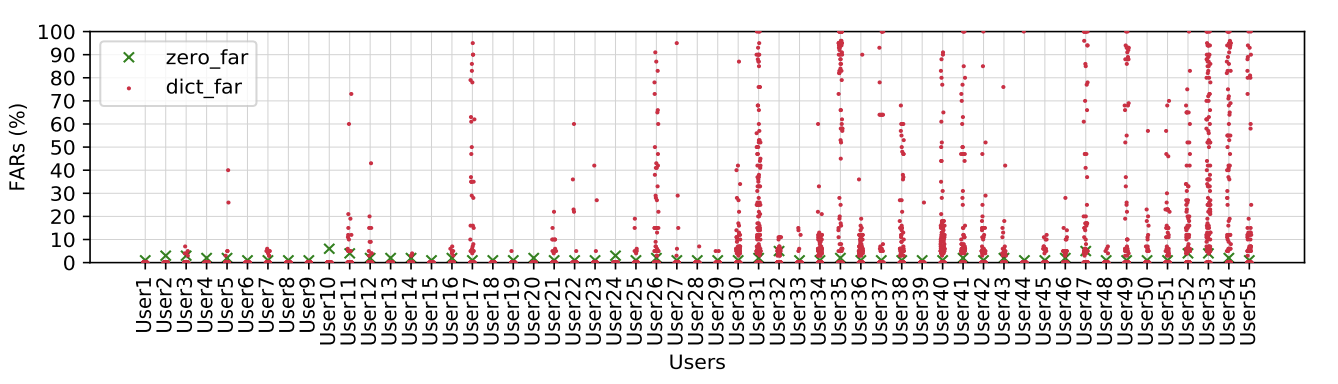}
    \caption{The distribution of \textit{FAR}s under Zero-effort attack (\textit{zero\_fars}) and \textit{FAR}s under Dictionary-effort attack (\textit{dict\_fars}) across genuine user population for (a+g+m+r)-based models respectively. $178$ \textit{IMUGait} patterns were tried for each genuine user.}
    \label{uservsuniqueattacks_all_svm}
\end{figure*}

We recorded the False Accept Rate (FAR) for the dictionary entry that caused the maximum damage. A deeper investigation revealed that multiple dictionary entries caused significant damage for some sets of users (see Figures \ref{uservsuniqueattacks_a_svm} and \ref{uservsuniqueattacks_all_svm}). In contrast, some sets of users remained unaffected as none of the dictionary entries could expand false accept rates for those users. We anticipated such a possibility because our dictionary is limited to nine users and has a total of $178$ settings only.

\section{Results and discussion}
\label{ResultsDiscussion}
\subsection{Sensor and classifier level error analysis} The \textit{FAR} heatmaps presented in Figure \ref{dict-effort-sensor-level} suggest that the Dictionary attack increased the average \textit{FAR} substantially (on an average from 6\% to 59\%). The average \textit{FAR} of accelerometer-based models (the best performing individual sensor under the Zero-effort attack scenario) increased from 5\% to 56\%. A similar trend was observed for the fusion-based models as the average \textit{FAR} of (a+g+m), (a+g+r), and (a+g+m+r)-based models reached 56\%, 53\%, and 52\% from 3\%. Figure \ref{dict-effort-sensor-level} suggests HTER increase from 10\% to 37\%. An authentication system with 37\% HTER is too weak to be useful for any real-world application scenario.  

Different degree of attack impact was observed for different implementations of the authentication system. For example, the \textit{SVM}-based models were the least affected ones as the \textit{FAR} increased from 5\% to 53\% followed by \textit{MLP} from 3\% to 56\%, \textit{LReg} from 7\% to 62\%, \textit{RanFor} from 9\% to 62\%, and \textit{kNN} from 4\% to 64\%. This was unsurprising as SVM specifically focuses on maximizing the distance between the support vectors. Both FAR and HTER heatmaps suggest that SVM-based implementations showed the maximum resilience to the Dictionary attack followed by \textit{kNN}, \textit{LReg}, \textit{MLP} and \textit{RanFor}. 

From the heatmaps presented in Figure \ref{dict-effort-sensor-level}, we can safely conclude that the Dictionary-based attack was alarmingly successful. The error heatmaps presented in Figure \ref{dict-effort-sensor-level} showed the overall performance of the Dictionary-based circumvention; however, they fail to reveal the attack's impact on individual user authentication models. Therefore, we present the user-level error analysis in the following sections. We restrict our experimentation to the two most resilient implementations of the attack environment, one from the individual sensor category and the other from the fused sensor category. 

\subsection{User level analysis} This section analyses the impact of the dictionary attack on the individual user authentication model. The FAR heatmaps in Figure \ref{dict-effort-cls-user-a} suggest that 11 users \textit{(User2, User3, User9, User15, User16, User19, User20, User21, User22, User23, and User29)} showed excellent resilience to the Dictionary attack. In comparison, the attack severely impacted the rest 44 users. A deeper analysis suggests that 80\% of the total accelerometer-based user models accepted more than 50\% impostors under Dictionary-attack. 

On the other hand, the FAR heatmaps in Figure \ref{dict-effort-cls-user-all} suggest that the fusion-based models showed more resilience to the Dictionary attack as 18 (User1, User2, User3, User4, User6, User7, User8, User9, User12, User13, User14, User15, User16, User18, User19, User20, User28, and User29) of the 55 users remained unaffected (the FAR remained below the baseline). The rest (67\%) of the fusion-based authentication models were severely ($\ge$ 50\% FAR) impacted by the Dictionary attack. 

From this, we conclude that the Dictionary attack is highly effective against individual sensors and fusion-based implementations for many users. At the same time, we observed that the attack did not work on some users. The explanation for this observation lies in the biometric menagerie defined by \cite{LambsAndWolves1998} and \cite{TheBiometricMenagerieExtended}. The most impacted users belong to the \textit{Lambs} class, and the least impacted users to the \textit{Doves} class \cite{TheBiometricMenagerieExtended}. Although time-consuming, the systematic procedure presented in Section \ref{DictionaryDataset} for dictionary creation can be easily adapted to add more imitators to increase the attack's success. 

Further investigation into the demographics of genuine and imitators revealed that the demographic overlap for (1) users with the lowest error rates and imitators and (2) users with the highest error rates and imitators were quite similar. That suggests that the imitators overlapping with the target do not necessarily have any advantage over those who do not overlap with the target users. 

\subsection{Probability of finding a matching gait password from the dictionary?} The previous result plots emphasized the analysis of the error rates obtained by the entry that achieved the highest increase in the FAR. We were curious to know how other entries in the dictionary fare against the user authentication models. What was their success rate? Therefore, we report \textit{FAR}s achieved by each \textit{IMUGait} entry of the dictionary against the most resilient implementations. Following previous choices, we present and analyze the error rates of \textit{SVM}-based models. Among \textit{SVM}-based models, we chose the accelerometer under the individual sensor category and (a+g+m+r) under the fused-sensor category. Figure \ref{uservsuniqueattacks_a_svm} presents the \textit{FAR}s under Zero- and Dictionary-effort circumvention scenarios for accelerometer-based authentication models. Similarly, Figure \ref{uservsuniqueattacks_all_svm} depicts the \textit{FAR}s obtained under Zero- and Dictionary-effort circumvention scenarios for (a+g+m+r)-based authentication model.  

We can see that more than one \textit{IMUGait} pattern in the dictionary has succeeded in increasing the \textit{FAR} beyond $zero\_fars$ as marked with green cross (x) symbol (see User4, User8, User9, User13, User27, User28, User11, User15, User39, User48, and User7). Moreover, the \textit{FAR}s obtained by some \textit{IMUGait} pattern from the dictionary are alarmingly high. On the other hand, for some users (User2, User3, User16, User19, User20, User21, User22, User23, User24, User25, and User29), none of the \textit{IMUGait} patterns from the dictionary were able to obtain higher \textit{FAR}s than $zero\_far$. In other words, 11 of the 55 users remained unaffected in the dictionary-effort circumvention environment. The reason behind this is the small number of entries in the dictionary. We posit that much more diverse entries in the dictionary would make the attack more effective and increase the likelihood of finding a highly damaging entry in the dictionary for every user. 

\subsection{Impact beyond \textit{IMUGait}} Although this work focuses on \textit{IMUGait}, the proposed methods can be extended to other behavioral biometrics. For example, one could (1) define several factors that dictate the characteristics of human swipe gestures, (2) determine several levels of these factors, (3) find a relationship between the factors and the feature extracted from raw sensor readings, and (4) use the relationship to train a human or robot to produce swipe gestures that will be close to that of the target individual. The proposed technique can also train robots to mimic well-defined human behavior. 

\subsection{Possible countermeasures} Possible countermeasures for the presented attack method could include the fusion of sensor readings collected from different devices, e.g., smartphone, smartwatch, and smart ring \cite{ZEMFANitesh}. However, that would mean everyone needs to have multiple devices. On the other hand, we aim to investigate if the use of deep generative models for generating data and using the same retrain the model would help classifiers sketch a more robust boundary between the genuine and impostor samples \cite{mo2022ictganan, TrainingTestingUsingOtherUsers, IJCB2021GANTouch, GANTouchTBIOM}.

\section{Conclusion and future work}
\label{Conclusion}
This paper introduced a new adversarial approach to exploit vulnerabilities in gait authentication systems, specifically those using smartphones' Inertial Measurement Unit (IMU). Inspired by dictionary attacks on traditional systems like PINs or passwords, the research explored the possibility of building a dictionary of IMUGait patterns. Nine physically diverse imitators created a dictionary of 178 unique IMUGait patterns. User authentication models created from 55 genuine users's data were tested against the dictionary, revealing significant weaknesses in IMUGait-based systems. The results challenge the widely held belief that such biometric systems are inherently secure, emphasizing the necessity for further research and enhanced security measures.

In the future, we will focus on incorporating a broader range of datasets, testing deep learning-based authentication models' resilience, developing countermeasures for the proposed attack, and expanding the idea to test the robustness of behavioral biometrics beyond IMU-gait. 
 
\bibliographystyle{ACM-Reference-Format}
\balance
\bibliography{references}
\end{document}